\documentclass[twocolumn,showpacs,amsmath,amssymb]{revtex4}
\usepackage{graphicx}
\usepackage{dcolumn}
\usepackage{bm}
\newcommand{\gb}{{\cal O}}
\newcommand{\BR}{{\cal B}}

\newcommand{\jpc}{J^{PC}}

\newcommand{\pip}{\pi^+}
\newcommand{\pim}{\pi^-}
\newcommand{\piz}{\pi^0}

\newcommand{\etap}{\eta^{\prime}}
\newcommand{\omegap}{\omega^{\prime}}
\newcommand{\etac}{\eta_c}
\newcommand{\etacp}{\eta_c^{\prime}}

\newcommand{\chicj}{\chi_{cJ}}
\newcommand{\psp}{\psi^{\prime}}
\newcommand{\psip}{\psi(2S)}
\newcommand{\pspp}{\psi^{\prime \prime}}

\newcommand{\jpsi}{J/\psi}

\newcommand{\EE}{e^+e^-}
\newcommand{\MM}{\mu^+\mu^-}
\newcommand{\TT}{\tau^+\tau^-}

\newcommand{\pp}{\pi^+\pi^-}
\newcommand{\kk}{K^+K^-}
\newcommand{\kskl}{K^0_SK^0_L}
\newcommand{\kstk}{K^* \overline{K}}
\newcommand{\KKSC}{K^{*+}K^-}
\newcommand{\KKSN}{K^{*0}\overline{K^0}}

\newcommand{\rpi}{\rho\pi}
\newcommand{\OP}{\omega\pi^0}
\newcommand{\RP}{\rho\pi}

\newcommand{\rhopi}{\rho\pi}

\newcommand{\ccbar}{c\bar{c}}
\newcommand{\ccb}{c\overline{c}}

\newcommand{\qqb}{q\overline{q}}

\newcommand{\BBb}{B\overline{B}}
\newcommand{\KKb}{K\overline{K}}
\newcommand{\ppb}{p\overline{p}}

\newcommand{\ddb}{D\overline{D}}
\newcommand{\DDb}{D\overline{D}}

\newcommand{\VP}{1^-0^-}
\newcommand{\PP}{0^-0^-}
\newcommand{\jpsipp}{J/\psi \pi^+\pi^-}
\newcommand{\PPJP}{\pi^+\pi^- J/\psi}
\newcommand{\ra}{\rightarrow}
\newcommand{\beq}{\begin{equation}}
\newcommand{\eeq}{\end{equation}}
\newcommand{\bfg}{\begin{figure}}
\newcommand{\efg}{\end{figure}}
\newcommand{\btbl}{\begin{table}}
\newcommand{\etbl}{\end{table}}
\newcommand{\btbu}{\begin{tabular}}
\newcommand{\etbu}{\end{tabular}}
\def\eref#1{(\ref{#1})}
\def\Journal#1#2#3#4{{#1} {\bf #2}, #3 (#4)}
\def\CTP{Commun. Theor. Phys.}
\def\IJMPA{Int. J. Mod. Phys. A}

\def\NPA{Nucl. Phys. A}
\def\PL{Phys. Lett. }
\def\PLB{Phys. Lett. B}

\def\PRL{Phys. Rev. Lett.}
\def\PRD{Phys. Rev. D}
\def\PRC{Phys. Rev. C}
\def\PRP{Phys. Rep.}
\def\PTPS{Prog. Theor. Phys. Suppl.}

\begin{document}
\preprint{Draft-PRD}

\title{Study of the $\RP$ Puzzle in Charmonium Decays}
\author{MO Xiao-Hu}
\email{moxh@mail.ihep.ac.cn}
\author{YUAN Chang-Zheng}
\email{yuancz@mail.ihep.ac.cn}
\author{WANG Ping}
\email{wangp@mail.ihep.ac.cn}
\affiliation{Institute of High Energy Physics, CAS, Beijing
100039, China}

\date{\today}

\begin{abstract}
The theoretical explanations about the ``$\RP$ puzzle'' in charmonium decays 
are reviewed extensively, and the comparison of theoretical predications 
with experimental data is made whenever possible. 
Three methods to estimate the ratio of the branching fractions of 
$\jpsi$ and $\psp$ decays are also discussed. It is pointed out that 
in order to understand the $\RP$ puzzle, and the dynamics of charmonium decays, 
systematic studies should be made in theory, phenomenology and experiment aspects.

\end{abstract}


\maketitle

\section{Introduction}

Crisply defined experimental puzzles in high-energy physics always
have the prospect of leading to new discoveries. One prominent
example is the $\theta$-$\tau$ puzzle of 1956 which led to the parity
revolution. Therefore puzzles in physics often draw considerable 
attention to theorists. 
The ratios of hadronic decays of the $\psi(3686)$ (shortened as
$\psp$) to the same decays of the $\jpsi$ 
is such a puzzle which has been studied substantially since 1983.

Since the OZI suppressed decays of $\jpsi$ and $\psp$ to hadrons 
are via three gluons or a photon, in either case, the perturbative QCD 
(pQCD) provides a relation~\cite{appelquist}
\begin{eqnarray}
Q_h &=&\frac{{\cal B}_{\psp \ra h}}{{\cal B}_{\jpsi \ra h}}
=\frac{{\cal B}_{\psp \ra \EE}}{{\cal B}_{\jpsi \ra \EE}} 
\approx 12.7\%~. \label{qcdrule}
\end{eqnarray}
This relation is referred to as the ``12\% rule'' which is expected 
to be held to a reasonable good degree for both inclusive and 
exclusive decays. The so-called ``$\RP$ puzzle'' is that the 
prediction by Eq.~\eref{qcdrule} is severely violated in the $\RP$ 
and several other decay channels. The first evidence for this 
effect was found by Mark-II Collaboration in 1983~\cite{mk2}. 
From then on many theoretical explanations have been put forth 
to decipher this puzzle. 

With the recent experiment results from BESII and CLEOc about 
$\jpsi$ and $\psp$ two-body decays, such as vector-pseudoscalar ($VP$), 
vector-tensor ($VT$), pseudoscalar-pseudoscalar ($PP$), and 
baryon-antibaryon ($\BBb$) modes, and about multi-body decays at 
the $\jpsi$, the $\psp$ or even at the $\psi(3770)$ 
(shortened as $\pspp$)~\cite{bes3pi}-\cite{cleoklks2}, 
a variety of solutions proposed for the puzzle can be tested 
at the level of higher accuracy. 
In this treatise, we survey the theoretical works on the $\RP$ puzzle 
and compare them with the available experimental data.  
From the theoretical point of view,
since the $Q$-value is smaller than 12\% for $\rpi$, it may be caused 
either by enhanced or suppressed $\jpsi$ decay rate.
Another possibility is by both. So we classify the relevant theoretical 
speculations into three categories:

\begin{enumerate}
\item $\jpsi$-enhancement hypothesis, which attributes the small 
$Q$-value to the enhanced branching fraction of $\jpsi$ decays.
\item $\psp$-suppress hypothesis, which attributes the small $Q$-value 
to the suppressed branching fraction of $\psp$ decays.
\item Other hypotheses, which are not included in the 
above two categories.
\end{enumerate}

In the following content, first reviewed are the theoretical works 
on $\RP$ puzzle, and the predictions from them are compared with the newly available  
experimental results; then expounded are three methods to estimate 
the ratio between $\psp$ and $\jpsi$ decays into the same final states;
after that some comments are made on the implications from the review 
on $\RP$ puzzle; at last there is a short summary.

\section{Review of theoretical works on $\RP$ puzzle}\label{sect_rvrpi}

\subsection{$\jpsi$-enhancement Theory}
In the earlier days of the $\rpi$ puzzle, it was noticed that
the decay of $1^{--}$ charmonium into $\rpi$ final state violates the 
Hadronic Helicity Conservation (HHC) theorem (see below for expound)~\cite{brodsky81}, 
and so such decay should be suppressed. Therefore people think there must 
be some mechanism which leads to the great enhancement for $\jpsi \to \rpi$ decays. 
The two schemes presented in this section were proposed following this line of reasoning.

\subsubsection{$\jpsi$-glueball Admixture Scheme}

The idea of $\jpsi$ decays via a glueball was proposed by 
Freund and Nambu~\cite{freund} (FN hereafter) soon after the discovery of 
$\jpsi$ particle to explain the breaking of Okubo-Zweig-Iizuka (OZI) 
rule~\cite{ozi}. In such mechanism, the breaking results from the mixing 
of the $\omega$, $\phi$, and $\jpsi$ mesons with an $SU(4)$-singlet 
vector meson $\gb$. They found that such an $\gb$  meson should lie in 
the 1.4-1.8 GeV/$c^2$ mass range with the width greater than 40 MeV/$c^2$, 
and it should decay copiously into $\RP$, $\kstk$ while exhibiting 
severe suppression of decays into $\KKb$, $\EE$ and $\MM$ modes. 
These authors presented several quantitative predications for 
experimental search. Two of them are
$$ R_1 =
\frac{\Gamma_{\jpsi \to \RP}}{\Gamma_{\phi \to \RP}} =0.0115-0.087~,$$
$$ R_2 =
\frac{\Gamma_{\jpsi\to\KKb}}{\Gamma_{\jpsi\to\RP}} < 8\times 10^{-5}~.$$
With the current available data and using the three pions final state as 
a substitute for $\RP$ in both $\phi$~\cite{pdg06} and 
$\jpsi$~\cite{bes3pi,babar3pi} decays, we obtain the first ratio 
$R_1 \approx 0.003$, which is almost one order of magnitude 
smaller than the predication. For the second ratio, by virtue of 
PDG~\cite{pdg06} value for $\kk$ and new experiment result for 
$\kskl$~\cite{jpsikskl}, it is estimated that $\BR(\jpsi\to\KKb)
\sim 10^{-4}$, together with the results for 
$\RP$~\cite{bes3pi,babar3pi}, we have $R_2 \sim 10^{-2}$ which is much
larger than the predication. 

The first attempt to explain the $\RP$ puzzle in terms of a glueball near 
$\jpsi$ was proposed by Hou and Soni~\cite{houws} (HS hereafter). 
They attributed the enhancement of $\jpsi \to \kstk$ and $\jpsi \to \RP$ 
decay modes to the mixing of the $\psi$ with a $\jpc=1^{--}$ vector gluonium, 
also designated by $\gb$. The differences between FN's and HS's pictures 
lie in the following aspects:
\begin{itemize}
\item Based on the potential model applied to glueball, the mass of 
a low-lying three-glue state is estimated to be around 2.4 GeV/$c^2$~\cite{houwsa}, 
rather than 1.4 to 1.8 GeV/$c^2$ in Ref.~\cite{freund}.
\item The mixing of $\gb$ with $\psp$ is taken into account, which has
been ignored in previous work.
\item Since the gauge coupling constant in QCD is momentum dependent, the
mixing parameter is taken to be a function of the invariant mass $q^2$,
which decreases rather sharply with the increase in $q^2$. Such propagator
effect gives rise to a more suppression on the decay rates of $\psp$ relative
to $\jpsi$ for decays to $\RP$ and $\kstk$ channels.
\end{itemize}

By virtue of their assumption, HS suggested a search for the vector
gluonium state in certain hadronic decays of the $\psp$, such as $\psp \to \pi\pi +X$,
$\eta +X$, $\etap +X$, where $X$ decays into VP final states~\cite{houws}.

Based on HS's idea, Brodsky, Lepage, and Tuan~\cite{brodsky87} 
(BLT) refined the glueball hypothesis for the $\RP$ puzzle. They assumed 
the general validity of the pQCD theorem that the total hadron helicity is 
conserved in high-momentum-transfer exclusive processes, in which case 
the decays to $\rpi$ and $\kstk$ are forbidden for both the $\jpsi$ and $\psp$. 
This pQCD theorem is often referred to as the rule of Hadronic Helicity 
Conservation (HHC)~\cite{brodsky81}, which is based on the assumption of 
the short-range ``pointlike'' interactions among the constituent quarks
throughout. For instance, $\jpsi(\ccbar) \to 3g$ has a short range 
$\simeq 1/m_c$ associated with the short time scale of interaction.
Nevertheless, if subsequently the three gluons were to resonate forming
an intermediate gluonium state $\gb$ which has a large transverse size 
covering an extended time period, then HHC would become invalid. 
In essence the HS model takes over in this latter stage. 

Final states $h$ which proceed only through the intermediate 
gluonium state satisfy the ratio
\beq
Q_h = \frac{\BR(\psp \to \EE)}{\BR(\jpsi \to \EE)}
      \frac{(M_{\jpsi}- M_{\gb})^2+\Gamma^2_{\gb}/4}
           {(M_{\psp}- M_{\gb})^2+\Gamma^2_{\gb}/4}~.
\label{brodskyratio}
\eeq
The $Q_h$ is small if the mass of $\gb$ is close to the mass of $\jpsi$. 
The experimental limits at that time~\cite{pdg86,mk2,brodsky87} imply that 
the $\gb$ mass is within 80 MeV/$c^2$ of the mass of $\jpsi$ and its
total width is less than 160 MeV/$c^2$. Brodsky {\em et al.} recommended 
a direct way to search for $\gb$, that is to scan the $\EE \to VP$
cross section across the $\jpsi$ resonance.

  Another related work by Chan and Hou~\cite{chan} studied
the mixing angle $\theta_{{\cal O}\psi}$ and the mixing amplitude
$f_{{\cal O}\psi}$ of the $\jpsi$ and vector glueball ${\cal O}$ based on
the framework of potential models of heavy quarks and constituent gluons.
They obtain $|\tan \theta_{{\cal O}\psi}|=0.015$ and 
$f_{{\cal O}\psi}(m^2_{{\cal O}\psi})=0.008 \mbox{ GeV}^2$. 

On the experimental part, BES has searched for this hypothetical particle 
in a $\rpi$ scan across the $\jpsi$ mass region in $\EE$ annihilations 
as well as in the decays of $\psp \to \pi\pi \gb$, $\gb \to \rpi$, and 
found no evidence for its existence~\cite{besvp,xyh}. The data constrains 
the mass and width of the $\gb$ to the range $\mid M_{\gb}-M_{\jpsi}\mid < 80$ 
MeV/$c^2$ and 4 $< \Gamma_{\gb} < 50 $ MeV/$c^2$~\cite{harris}. 
Although the absence of distortion in BES energy scan of $\jpsi \to \rpi$ 
does not rule out $M_{\gb} \simeq M_{\jpsi}$, it puts a lower bound to 
$\Gamma_{\gb}$. However, as indicated in Ref.~\cite{chen}, the 
experimentally constrained mass is several hundred MeV/$c^2$ lower than 
the mass of the lightest vector glueball calculated in lattice simulations 
of QCD without dynamical quarks~\cite{peardon}. 

Recently, the experimental data from  BES and CLEOc turned out to be 
unfavorable to this glueball hypothesis. Among them is the observed 
large branching fractions of the isospin-violating VP mode 
$\psp \to \omega\pi^{0}$~\cite{besvp1,besvp2,cleocvp}. This 
contradicts with the assertion that the pattern of suppression 
depends on the spin-parity of the final state mesons. 
In addition, according to BLT's analysis, one obtains the relation~\cite{suzukia}
$$ \frac{\BR(\jpsi \to \OP)}{\BR(\jpsi \to \rho^0 \pi^0)}<0.0037~$$
which is much smaller than the PDG06 value 0.08~\cite{pdg06}. 
Another experimental result which is unfavorable to such hypothesis 
is the suppression of $\psp$ decays into vector-tensor (VT)
final states~\cite{bes1vt,bes2vt}. 
Since hadronic VT decays, unlike the VP decays, conserve HHC, some other
mechanisms must be responsible for this suppression in the model.
Furthermore, it has been argued that 
the $\gb$ may also explain the decay of $J/\psi$ into $\phi f_{0}$ 
(named previously $S^{\ast}$) but not to $\rho a_0(980)$ 
(named previously $\delta$), since the $\gb$ mixes with the $\phi$ and 
enhances a mode that would be otherwise suppressed~\cite{brodsky87}. 
However, the decay $\psp\to\phi f_{0}$~\cite{harris} is not suppressed 
as found by experiments, which implies the absence of anomalous 
enhancement in $J/\psi\rightarrow\phi f_{0}$, thus contradicts with 
this explanation. Anselmino {\em et al.} extended the idea of 
$\jpsi$-$\gb$ mixing to the case of $\eta_c\rightarrow VV$ and 
$p\overline{p}$~\cite{anselm}. They suggested that the enhancement of
these decays can be attributed to the presence of a trigluonium 
pseudoscalar state with a mass close to the $\eta_c$ mass. So far 
there is no experimental evidence for such a state.

In fact, as pointed out in Ref.~\cite{chaichian}, this glueball 
explanation has some unanswered questions:
(i) Why only the $\rpi$ and $\kstk$ channels are affected but not 
$\gb \to 5 \pi$ etc. viz., one must assume in an {\em ad hoc} way  
that the $\gb$ couples predominantly to $\rpi$ and $\kstk$; and (ii) 
if such a narrow heavy $1^{--}$ gluonium state exists, 
why have the narrow $0^{++}$, $2^{++}$ states 
not been found, which should be lighter and easier to search?

\subsubsection{Intrinsic-charm-component Scheme}

 Brodsky and Karliner (BK) put forth an explanation for the puzzle 
based on the existence of the intrinsic charm $|\qqb \ccb \rangle$ Fock
components of the light vector mesons~\cite{brokar}. They noticed
the fact that quantum fluctuations in a QCD bound state wave function 
will inevitably produce Fock states containing heavy quark pairs. 
The intrinsic heavy quark pairs are multiconnected to the valence quarks 
of the light hadrons, and the wave functions describing these 
configurations will have maximal amplitude at minimal off-shellness and 
minimal invariant mass. In the case of the $\rho$ meson, with the 
consideration of the light-cone Fock representation:
$$ \rho^+ = \psi^{\rho}_{u\overline{d}} |u\overline{d} \rangle +
\psi^{\rho}_{u\overline{d}\ccb} |u\overline{d}\ccb \rangle + \cdots~.$$
Here we expect the wave function of the $\ccb$ quarks to be in an $S$-wave
configuration with no nodes in its radial dependence, in order to minimize
the kinetic energy of the charm quarks and thus also minimize the total
invariant mass.

The presence of the $|u\overline{d}\ccb \rangle$ Fock state in the $\rho$
allows the $\jpsi \to \rpi$ decay to proceed through rearrangement of 
the incoming and outgoing quark lines; in fact, the $|u\overline{d}\ccb 
\rangle$ Fock state wave function has a good overlap with the radial and
spin $|\ccb \rangle$ and $|u\overline{d} \rangle$ wave functions of the
$\jpsi$ and pion. On the other hand, the overlap with the $\psp$ will be
suppressed, since the radial wave function of the $n=2$ quarkonium state
is orthogonal to the nodeless $\ccb$ in the $|u\overline{d}\ccb \rangle$
state of the $\rpi$. Similarly, the $|u\overline{s}\ccb \rangle$ Fock
component of the $K^*$ favors the $\jpsi K$ configuration, 
allowing the $\jpsi \to \kstk$ decay to also proceed by quark line 
rearrangement, rather than $\ccb$ annihilation.

These authors also suggested comparing branching fractions 
for the $\etac$ and $\etacp$ decays 
as clues to the importance of $\eta_c$ intrinsic charm excitations in 
the wavefunctions of light hadrons.

\subsection{$\psp$-suppress Theory}
The hypothesis of the existence of a glueball to explain the $\rpi$
has been questioned soon after it was proposed. 
In addition, it is also pointed out~\cite{Chernyak99} that the helicity 
suppression is not a strong constraint in the charmonium energy scale. 
Under such case, one comes naturally to the idea that it is not 
$\jpsi \to \rpi$ which is enhanced, but rather $\psp \to \rpi$ 
which is suppressed. Seven explanations or models collected in this section 
are put forth along this line.

\subsubsection{Sequential-fragmentation Model}

 Karl and Roberts have suggested explain the $\rpi$ puzzle based on 
the mechanism of sequential quark pair creation~\cite{karl}. The 
idea is that the quark-antiquark pairs are produced sequentially,
as a result the amplitude to produce two mesons in their ground
state is an oscillatory function of the total energy of the system.
They argue that the oscillatory fragmentation probability could have
a minimum near the mass of $\psp$, which provides an explanation for
the suppressed $\psp$ decay. Even though their evaluations could 
generally accommodate the data for decays of $\jpsi$ and $\psp$ to 
$\rpi$ and $K^{\ast}\overline{K}$, it runs into difficulties when 
it is extrapolated to $\Upsilon$ decays. According to their calculation, 
the oscillations of probability amplitude are damped out in the region 
of the $\Upsilon$ resonances, so the $\rpi$ channel is present in 
the decay of all $\Upsilon$, $\Upsilon^{\prime}$, $\Upsilon^{\prime\prime}$, 
$\cdots$ resonances with a common rate. This leads to a prediction 
$\Gamma (\Upsilon \to \rpi)=0.05$ keV, or equivalently 
$\BR(\Upsilon \to \rpi)=9.4 \times 10^{-4}$, which is above
the current upper limit $\BR(\Upsilon \to \rpi)<2 \times 
10^{-4}$~\cite{pdg06}. Moreover, their calculation seems hard to  
explain the large branching fraction for $\phi$ decays to 
$\rpi$~\cite{pdg06} due to the fact that their fragmentation 
probability tends to zero as the mass of the  $\rpi$ decaying system 
approaches 1GeV/$c^2$.

 In a further analysis~\cite{karla}, Karl and Tuan pointed out that if 
a suppression is observed in three-meson channels the explanation 
based on sequential pair creation would be undermined. Recently such
a suppressed channel, viz. $\phi KK$, is found by CLEOc~\cite{cleolhd1}. 

\subsubsection{Exponential-form-factor Model}

  Guided by suppressed ratios of $\psp$ to $\jpsi$ decays to two-body 
hadronic modes, Chaichian and T\"{o}rnqvist suggested~\cite{chaichian} 
that the hadronic form factors fall exponentially as described by the 
overlap of wave functions within a nonrelativistic quark model. This 
behavior explains the drastically suppressed two-body decay rates of 
the $\psp$ compared with those of the $\jpsi$. Recently, the report on 
observation of a number of VP channels in 
$\psp$ decays~\cite{besvp1,besvp2,cleocvp}
such as $\omega \etap$, $\phi \etap$, $\rho \etap$ has proved that
the predicted decay fractions are overestimated. Moreover, 
the branching fraction for $\OP$~\cite{pdg06},
is well below the prediction by this model which is $1.04\times 10^{-4}$.

 Another problem of the model is that it does not single out just
the VP channel, the other channels, for example VT channel, are also 
estimated to have small branching fractions which are not compatible with 
the BES measured results~\cite{bes2vt}.

\subsubsection{Generalized Hindered M1 Transition Model}
 
  A so-called generalized hindered M1 transition model is proposed by 
Pinsky as a solution for the puzzle~\cite{pinsky}. It is argued that 
because $\jpsi \to \gamma\eta$ is an allowed M1 transition while 
$\psp \to \gamma\etap$ is hindered (in the nonrelativistic limit), using
the vector-dominance model to relate $\psp \to \gamma\etap$ to 
$\psp \to \psi\etap$ one could find the coupling $G_{\psp \psi \etac}$
is much smaller than $G_{\psi \psi \etac}$, and then by analogy, the
coupling $G_{\omegap \rho \pi}$ would be much smaller than 
$G_{\omega \rho \pi}$. Here $G_{\omega \rho \pi}$ can be extracted from 
data by virtue of the analysis using the vector-dominance model and a 
standard parameterization of OZI process~\cite{pinskya}. Then assuming 
$\psp\to\rpi$ to proceed via $\psp$-$\omegap$ mixing, while $\jpsi\to\rpi$
via $\psi$-$\omega$ mixing, one would find that $\psp\to\rpi$ is much more 
severely suppressed than $\jpsi\to\rpi$. The similar estimation could be
preformed for $\kstk$ and other VP final state, and one can expect a 
suppressed $Q$:
\beq
\frac{\BR(\psp\to VP)}{\BR(\psi\to VP)}=
1.47 \frac{\Gamma_{tot}(\psi)}{\Gamma_{tot}(\psp)}
     \left( \frac{G_{V^{\prime}VP}}{G_{VVP}} \right)^2 
     \frac{F_{V^{\prime}}}{F_{V}}=0.06\%~,
\label{pinskyqval}
\eeq
where $F_{V^{\prime}}/F_{V}=0.3$, $G_{\omegap\rho\pi}/G_{\omega\rho\pi}
=0.066$ according to Ref.~\cite{pinsky}. This $Q$ is much smaller than
the present experimental results~\cite{besvp1,besvp2,cleocvp}. 
 
  Moreover, in this model, the coupling $G_{\omegap\omega f_2}$ for 
$\omegap\to \omega f_2$ should not be suppressed because by analogy the 
coupling $G_{\psp\psi \chi_{c2}}$ is not small due to the fact that the E1 
transition $\psp\to\gamma \chi_{c2}$ is not hindered~\cite{chao}. 
Therefore via $\psp$-$\omegap$ mixing the $\psp \to \omegap\to \omega f_2$ 
decay is expected to be not suppressed, which contradicts the BES 
result~\cite{bes2vt}.

\subsubsection{Higher-Fock-state Scheme}

 Chen and Braaten (CB) proposed an explanation~\cite{chen} for
the $\rho\pi$ puzzle, arguing that the decay $J/\psi\rightarrow\rho\pi$
is dominated by a Fock state in which the $c\overline{c}$ pair is
in a color-octet $^{3}S_{1}$ state which decays via
$c\overline{c}\rightarrow q\overline{q}$, while the suppression of
this decay mode for the $\psp$ is attributed to a
dynamical effect due to the small energy gap between the mass
of the $\psp$ and the $\ddb$ threshold. Using the BES data
on the branching fractions into $\rho\pi$ and $K^{\ast}\overline{K}$ as
input, they predicted the branching fractions for many other
VP decay modes of the $\psp$, as listed in Table~\ref{cb_preds},
from which we see most measured values falling in the scopes of 
predictions, but we also note for $\omega \pi$ mode, the deviation from
the prediction is obvious. Here it should be noticed that the
values adduced in Table~\ref{cb_preds} are calculated on the strength 
of the measured branching fractions from earlier experiment, the new 
measurements on the branching fractions for $\rpi$ and 
$K^{\ast 0} \overline{K}^{0}+c.c.$ from BES~\cite{besvp3,besrpi1} 
and CLEOc~\cite{cleocvp} may have impact on the corresponding evaluations. 
 
\btbl[htb]
\caption{\label{cb_preds}Predictions and measurements for $Q_{VP}$ in 
unit of 1\% for all VP final states. The value for $\rpi$ and 
$K^{\ast 0} \overline{K}^{0}+c.c.$ from Ref.~\cite{zhuys} were used as
input. The theoretical parameter $x=0.64$ is due to results of 
Ref.~\cite{bramon} and the experimental results come from 
Ref.~\cite{besvp1,besvp2,besvp3,yrhopi}.} \vskip 0.2 cm \center
\btbu{ccc} \hline \hline 
      VP        & $x=0.64$    &    Exp.            \\ \hline \hline 
   $\rpi$       & $0-0.25$    &  $ 0.13 \pm 0.03$  \\
$\KKSN +c.c.$   & $1.2-3.0$   &  $ 3.2  \pm 0.08$  \\
$\KKSC +c.c.$   & $0-0.36$    &  $ 0.59^{+0.27}_{-0.36}$ \\
$\omega \eta$   & $0-1.6$     &  $ < 2.0 $         \\
$\omega \etap$  & $12-55$     &  $ 19^{+15}_{-13}$ \\
$\phi \eta$     & $0.4-3.0$   &  $ 5.1  \pm 1.9 $  \\
$\phi \etap$    & $0.5-2.2$   &  $ 9.4  \pm 4.8 $  \\
$\rho \eta$     & $14-22$     &  $ 9.2^{+3.6}_{-3.3}$    \\
$\rho \etap$    & $12-20$     &  $ 17.8^{+15.9}_{-11.1}$ \\
$\omega \pi$    & $11-17$     &  $ 4.4^{+1.9}_{-1.6}$ \\ \hline \hline
\etbu
\etbl

   Besides the predictions in Table~\ref{cb_preds}, CB's proposal also has 
implications for the angular distributions for two-body decay modes. 
In general, the angular distribution has the form $1+\alpha \cos^2 \theta$, 
with $-1<\alpha<+1$. CB's conclusion implies that the parameter $\alpha$
for any two-body decay of the $\psp$ should be less than or equal
to that for the corresponding $\jpsi$ decays.
But this needs further supports from analyses based on large data 
sample in the future.

\subsubsection{Survival-chamonia-amplitude Explanation}

 A model put forward by G\'{e}rard and Weyers entertains the assumption 
that the three-gluon annihilation amplitude and the QED amplitude add 
incoherently in all channels in $\jpsi$ decays into light hadrons, while 
in the case of $\psp$ decays the dominant QCD annihilation amplitude is 
not into three gluons, but into a specific configuration of five 
gluons~\cite{gerard}. 
More precisely, they suggest that the strong annihilation
of the $\psp$ into light hadrons is a two-step process: in the first
step the $\psp$ goes into two gluons in a $0^{++}$ or $0^{-+}$ state and
an off-shell $h_c(3526)$; in the second step the off-shell $h_c$ annihilates
into three gluons to produce light hadrons. Their argument implies:
(a) to leading order there is no strong decay amplitude for the processes
$\psp \to \rpi$ and $\psp \to \kstk$; (b) the 12 \% rule should hold
for hadronic processes which take place via the QED amplitude only.
As far as the second implication is concerned, the present data give
different ratios between $\psp$ and $\jpsi$ decay for $\OP$ and $\pp$ final
states, both of which are electromagnetic processes. Here the event form
factor effect is taken into account~\cite{wymfofa}, the difference 
between two kinds of processes is still obvious. Besides the explanation 
for $\rpi$ puzzle, this model predicts a sizable $\psp \to  
(\pi^{+}\pi^{-} \mbox{ or } \eta)~h_{1}(1170)$ branching fraction.

  In a recent paper~\cite{artoisenet}, Artoisenet, G\'{e}rard and Weyers (AGW) 
update and sharpen the above idea which leads to a somewhat unconventional point
of view: all non-electromagnetic hadronic decays of the $\psp$ goes
via a transition amplitude which contain a $\ccb$ pair. AGW provide
two patterns for these two-step decays, the first is
\beq
\psp \to (2\mbox{NP}g)+(3g)~.
\label{eq_23dk}
\eeq
The physics picture is as follows: the excited $\ccb$ pair in the $\psp$
does not annihilate directly. Instead, it spits out two non-perturbative 
gluons $(2\mbox{NP}g)$ and survives in a lower $\ccb$ configuration 
($1^{--}$ or $1^{-+}$) which then eventually annihilate into $3g$. The
decays $\psp \to (2\pi)\jpsi$ and $\psp \to \eta\jpsi$ follow this 
pattern. The second pattern is
\beq
\psp \to (3\mbox{NP}g)+(2g)~,
\label{eq_32dk}
\eeq
where the lower $\ccb$ configuration ($0^{-+}$ or $0^{++}$) annihilates 
into $2g$. The only on-shell channel for this type of decays is
$\psp \to (3\pi)\etac$, whose branching fraction is estimated as $(1-2)\%$
level. Anyway, the recent measurement from CLEOc~\cite{cleorep} provides
the upper limit which is one order of magnitude below this theoretical 
prediction. 
Furthermore, the substitution of one photon for one gluon in 
Eqs.~\eref{eq_23dk} and \eref{eq_32dk} allows
\beq
\psp \to (2\mbox{NP}g)+(2g)+\gamma~.
\label{eq_22gdk}
\eeq
This pattern corresponds to on-shell radiative decays such as
$\psp \to (\pi^+ \pi^-) \etac \gamma$ and $\psp \to \eta \etac \gamma$,
which could be larger than the observed $\psp \to \etac \gamma$  mode.

Besides the above predications, AGW also estimate 
\begin{eqnarray}
\BR(\psp \to b_1 \eta)  &=& (1.3\pm 0.3)\times 10^{-3}~, \\
\BR(\psp \to h_1 \pi^0) &=& (1.9\pm 0.4)\times 10^{-3}~, \\
\BR(\jpsi\to b_1 \eta)  &\approx &
  \BR(\psp \to b_1 \eta) \approx 1 \%~. 
\end{eqnarray}
All these are to be tested by experiments.

\subsubsection{Nonvalence Component Explanation}
Since the $\psp$ is a highly excited state and close to the $\DDb$ 
threshold, it is suggested~\cite{Chernyak} that unlike the $\jpsi$, the
$\psp$ may be an admixture of large nonvalence components in its 
wave function. The so-called nonvalence component indicates the additional
gluon or the light quark-antiquark pair (or as in Ref.~\cite{Chernyak},
the $\ccb g$ component and a quasi-molecular $\DDb$ state), which makes
$\psp$ decays prominently distinctive from those of $\jpsi$. The
nonvalence component of the $\jpsi$ is expected to be less significant 
because it is the lowest state. In a following paper~\cite{Chernyak99}, 
Chernyak uses this picture to explain the $\rpi$ puzzle. He suggested that
the valence and nonvalence strong contributions interfere destructively
in $\rpi$ channel and consequently cancel to a large extent in the
total $\psp \to \rpi$ strong amplitude, while the role of nonvalence 
contributions is much less significant in $\jpsi \to \rpi$. From this
viewpoint, there is no deep reason for the experimentally observed 
very strong suppression of $\psp \to \rpi$, this is the result of a 
casual cancellation. 

Chernyak also tries to use the above idea to explain qualitatively 
other decay modes, such as $VT$, $AP$, $PP$, $VV$ and $\BBb$ decay.
However, such an idea remains a pure speculation,
and no concrete calculations have been performed up to now.

\subsubsection{$S$-$D$ Wave Mixing Scheme}

The $\pspp$ is viewed as a $D$-wave charmomium state. Although it is 
primarily $c\overline{c}(1^3D_1)$, its leptonic width indicates a 
contribution from mixing with $S$-wave states, mainly the nearby $\psi(2^3S_1)$. 
This leads Rosner to propose that the small $\rhopi$ branching fraction 
in $\psp$ decay is due to the cancellation of the $2S$ and $1D$ matrix elements.
By virtue of his scheme   
\beq
\begin{array}{l}
\langle\rhopi |\psp\rangle =\langle \rhopi | 2^3 S_1 \rangle \cos \theta
                  -\langle \rhopi | 1^3 D_1 \rangle \sin \theta~, \\
\langle\rhopi |\pspp\rangle=\langle \rhopi | 2^3 S_1 \rangle \sin \theta
                  +\langle \rhopi | 1^3 D_1 \rangle \cos \theta~,
\end{array}
\label{sdmix}
\eeq
where $\theta$ is the mixing angle between pure $\psi(2^3 S_1)$ and
$\psi(1^3D_1)$ states~\cite{ktchao} 
and is fitted from the leptonic widths of the $\pspp$ 
and the $\psp$ to be $(12 \pm 2)^{\circ}$~\cite{rosnersd}, which is 
consistent with the coupled channel estimates~\cite{eichten2,heikkila} and
with the ratio of $\psp$ and $\pspp$ partial widths to 
$\jpsipp$~\cite{kuang}. If the mixing and coupling of the $\psp$ and 
$\pspp$ lead to complete cancellation of $\psp \to \rpi$ decay 
($\langle\rhopi |\psp\rangle= 0$), the missing $\rpi$ decay mode of 
the $\psp$ shows up instead as decay mode of the $\pspp$, enhanced 
by the factor $1/\sin^2 \theta$, the concrete estimation shows 
that~\cite{rosnersd}
\beq
{\cal B}_{\pspp\rightarrow\rhopi}=(4.1\pm1.4)\times10^{-4}~~.
\label{brphi2}
\eeq 

To test this scenario by the data collected at the $\pspp$ in 
$\EE$ experiments, it has been pointed out~\cite{wymrpi2,wymfsvp} that 
the continuum contribution must be considered carefully. 
Specifically speaking, by Rosner's estimation, the Born order 
cross section for $\pspp \to \rpi$ is
\beq
\sigma_{\pspp \to \rpi}^{Born} = (4.8 \pm 1.9) \mbox{~pb}~,
\label{ressct}
\eeq 
which is comparable in magnitude to that of the continuum 
cross section, viz. 
\beq
\sigma^{Born}_{\EE \to \rpi}=4.4~\hbox{pb}~.
\label{consct}
\eeq
So what is observed is the coherent sum of the
two amplitudes. Based on the analysis of experimental data, 
it has been suggested that there be a universal phase between 
strong and electromagnetic amplitudes in charmonium decays. 
With this assumption, the strong decay amplitude interferes
with the continuum amplitude either maximum 
destructively, e.g. for $\rhopi$, $\omega \eta$ and $\KKSC$ or 
maximum constructively, e.g. for $\KKSN$.
The destructive interference leads to the phenomena that
the measured cross section on top of the resonance could be 
smaller than that off the resonance.  
Recent experimental results~\cite{cleolhd2,besrpi2} on $\rhopi$, 
$\omega \eta$ and $\KKSC$ have demonstrated such
interference pattern. This provides support to
the prediction of Eq.\eref{brphi2}.  However, 
to uniquely determine ${\cal B}_{\pspp\rightarrow\rhopi}$, 
current available experimental data are insufficient, 
the $\pspp$ resonance must be scanned~\cite{wymmcpspp}.
So the quantitative test of Rosner's scenario remains to 
be the task of the future experiments on $\tau$-Charm factories.

In the consequent study~\cite{psippkskl}, Wang, Yuan and Mo (WYM) 
extend the $S$-$D$ wave-mixing scenario to $PP$ final state, and give 
a constraint for $\pspp \to \kskl$ decay,
\beq
0.12 \pm 0.07 \le 10^{5}\times
\BR(\pspp \ra \kskl) \le 3.8 \pm 1.1~, 
\label{bound}
\eeq
which is compatible with both the BESII result
$\BR(\pspp \ra \kskl) < 2.1 \times 10^{-4}$ at 90\% 
C.L.~\cite{bespsppkskl} and the CLEOc result
$\BR(\pspp \ra \kskl) < 1.17 \times 10^{-5}$ 
at 90\% C.L.~\cite{cleoklks2}.
Extrapolating fore-study to all charmless decays~\cite{wymcmless},
WYM found that this scenario could accommodate large non-$\ddb$ decay
of the $\pspp$. Although the recent experimental measurement from 
CLEOc favors the zero non-$\ddb$ 
cross section at $\pspp$~\cite{cleondd},
the larger error could not rule out the existence of 
non-$\ddb$ branching fraction at 10\% level. 

\subsection{Other Explanations}
Besides the models in two previous sections, more speculations involving 
the $\rpi$ puzzle are collected in this section.

\subsubsection{Final State Interaction Scheme}

 Li, Bugg and Zou~\cite{lixq} (LBZ) pointed out that the final state 
interactions (FSI) in $\jpsi$ and $\psp$ decays give rise to effects 
which are of the same order as the tree level amplitudes, they argued
that $\jpsi \to \RP$ is strongly enhanced by the $a_2 \rho$ loop diagram,
while the direct tree-production for $\RP$ may be suppressed by the 
HHC mechanism~\cite{brodsky81}. The contribution
of the $a_1 \rho$ loop diagram is much smaller than that of $a_2 \rho$ 
loop for the $\jpsi \to \RP$, but they have similar strength for the
$\psp \to \RP$ and may cancel each other by interfering destructively.
The similar apparent suppression for $\psp \to \kstk$ and $f_2 \omega$
may also be explained by the $K^* \overline{K}^*_{2,1}$ and $b_1 \pi$
loop, respectively. Therefore, LBZ expected that FSI may provide a 
coherent explanation for all the observed suppressed modes of $\psp$ 
decays. However, as remarked in Ref.~\cite{tuan}, this interference model
does appear to have more assumptions that predictions. The only qualitative
prediction by LBZ is to check whether the $a_{1}\rho$ and 
$K_{1}^{\ast}\overline{K}^{\ast}$ production rates are large for the $\psp$.
So far, no such measurements have been reported. Nevertheless, useful 
information on $a_{1}\rho$ and $K_{1}^{\ast}\overline{K}^{\ast}$ could be 
obtained from the BES published data as shown in Refs.~\cite{bes0vt} and 
\cite{bes2vt}. The lack of evidence within the invariant mass distribution 
plots (see Fig.3 and Fig.5 of Ref.~\cite{bes0vt} or Ref.~\cite{bes2vt}) 
that the $\rpi$ recoiled against a $\rho$ for events of  
$\psp\to\rho^{0}\rho^{\pm}\pi^{\mp}$ and that $\pi^{\pm}K^{\mp}$ recoiled 
against a $K^{\ast 0}$ for events of $\psp\to\pi^{+}\pi^{-}K^{+}K^{-}$ 
suggests that they are unlikely to be the favored modes in $\psp$ decays.

\subsubsection{Large Phase Scheme}

  Suzuki gave another explanation based on FSI for $\jpsi$ 
decays~\cite{suzukia}. He performed a detailed amplitude analysis for 
$\jpsi \to \VP$ decay to test whether or not the short-distance FSI 
dominates over the long-distance FSI in the $\jpsi$ decay. His result 
indicates that there is a large phase between three-gluon and one-photon 
amplitudes. Since the large phase cannot be produced with the perturbative 
QCD interaction, the source of it must be in the long-distance part of 
strong interaction, namely, rescattering among hadrons in their inelastic 
energy region. Suzuki then performed the similar analysis for 
$\jpsi \to \PP$ decay, and obtained 
the same conclusion of the large phase~\cite{suzukib}. 
His analysis also shows that the exclusive decay rate at the $\jpsi$ is 
in line with that of the inclusive decay. This fact leads him to believe 
that the origin of the relative suppression of $\psp\to\VP$ to $\jpsi\to\VP$ 
is not in the $\jpsi$ but in the $\psp$. 

As to this large phase, 
in Ref.~\cite{gerard}, G\'{e}rard argued that this phase follows from 
the orthogonality of three-gluon and one-photon virtual processes. 
As a matter of fact, the conclusion of a large phase 
close to $90^\circ$ between  three-gluon and one-photon 
processes holds true for all two-body decays of $\jpsi$,
such as $1^+0^-$($90^\circ$)~\cite{suzukic},
$1^-0^-$ $(106 \pm 10)^\circ$~\cite{VPmode,castro}, 
$1^-1^-$ $(138 \pm 37)^\circ$~\cite{suzukib,castro,kopke},
$0^-0^-$ $(89.6 \pm 9.9)^\circ$~\cite{suzukib,castro,kopke} and
$N\overline{N}$ $(89 \pm 15)^\circ$~\cite{NNmode,castro}.

Nevertheless, as the $\psp$ decays were analyzed, the 
experimental data at first seemed to favor a phase
close to $180^\circ$~\cite{suzukic}, 
in contrary to the expectation that the decay of $\jpsi$ and $\psp$ 
should not be much different. 
However, as pointed out by Wang {\em et al.} that 
the previous published data did not take the 
continuum one-photon process into account. Their reanalysis
of the experimental data shows that   
the phase with 
value around $-90^{\circ}$ could fit $\psp\to\VP$ data~\cite{wymfsvp} 
and $\pm 90^{\circ}$ could fit $\psp\to\PP$ data~\cite{wymfspp}.
The latter is confirmed by more recent results by CLEO~\cite{cleocpp}. 
Furthermore, this large phase also shows in the OZI suppressed
decay modes of the $\pspp$. In many decays modes of the $\pspp$, 
the strong decay amplitudes have comparable strength as 
the non-resonance continuum amplitude, the large phase around $-90^\circ$ 
leads to destructive or constructive interference. Due to the destructive 
interference, the observed cross sections of some modes at the peak of 
the $\pspp$ are smaller than the cross section measured 
off-resonance~\cite{wymrpi2}. This is demonstrated by the data from 
CLEOc~\cite{cleolhd2}.

If the large phase between the
three-gluon and one-photon virtual processes
is universal not only in $\jpsi$ decays, but also in $\psp$ as well as
all charmonium or perhaps all quarkonium decays, then how this phase 
is related to the difference between the decays of $\jpsi$ 
and $\psp$ remains to be an unanswered question. 

\subsubsection{Mass Reduction Explanation}

  In the study~\cite{majp} of radiative decays of $1^{--}$ quarkonium 
into $\eta$ and $\etap$, Ma presented a QCD-factorization approach,
with which he obtained theoretical predictions in consistency with CLEOc
measurement. The largest possible uncertainties in analysis are from
the relativistic corrections for the value of the charm quark mass. Ma
argued that the effect of these uncertainties can be reduced by
using quarkonium masses instead of using quark mass. As an example
of such reduction, he provided a modified relation to the original 
12\% rule 
$$ Q_{\rpi} = \frac{\BR(\jpsi \to \rpi)}{\BR(\psp \to \rpi)}
            = \frac{M^8_{\jpsi}}{M^8_{\psp}} 
	      \frac{\BR(\jpsi \to \EE)}{\BR(\psp \to \EE)} $$
$$ = (3.6 \pm 0.6) \%~. $$
However, this value is much larger than the experimental result
given in Table~\ref{cb_preds}.

\subsubsection{Vector-meson-mixing Model}\label{sst_vmmm}

  Intending to give a comprehensive description of $\jpsi$ two-body 
decays, Clavelli and Intemann (CI) proposed a vector-meson-mixing model in 
which the vector mesons ($\rho,~\omega~,\phi~,\jpsi$) are regarded as 
being admixture of light-quark-antiquark state and charmed-quark-antiquark 
state~\cite{clavelli}. The coupling of the $\jpsi$ to any state of light 
quarks is then related to the corresponding coupling of the $\rho$, $\omega$, 
and $\phi$ to the same state. With few experiment inputs to determine 
the mixing parameters, CI calculated VP, PP, and BB decay rates for the 
$\jpsi$ as a function of the pseudoscalar mixing angle. Most of the 
predictions agree with the experiment results at the order of magnitude 
level, but discrepancy is obvious for some channels, such as $\kskl$ final 
state~\cite{jpsikskl}. CI also extended their model to the hadronic decays 
of the $\psp$. Nevertheless, their evaluations
for $\BR(\jpsi \to \OP)=3\times 10^{-5}$ and 
$\BR(\psp \to \OP)=3\times 10^{-3}$ contradict with the present 
results $(4.5 \pm 0.5) \times 10^{-4}$ and 
$(2.1 \pm 0.6)\times 10^{-4}$~\cite{pdg06}, respectively.

  Starting from effective Lagrangian whereby nonet-symmetry breaking
and pseudoscalar-meson mixing can be studied, Haber and Perrier 
parametrized the decay modes of
$\jpsi \to PP$ (also for $\jpsi \to VV$ or $\etac \to VP$),
$\jpsi \to VP$ (also for $\jpsi \to VT$ or $\etac \to VV$),
$\jpsi \to PPP$ (also for $\jpsi \to VVP$ or $\etac \to PPV$), and
$\etac \to PPP$ (also for $\jpsi \to PPV$ or $\etac \to VVP$)~\cite{haber}.
The experimental data are used to determine these phenomenological 
parameters. In a later work, Seiden, Sadrozinski and Haber took 
the doubly Okubo-Zweig-Iizuka suppression (DOZI) effect into consideration, 
and presented a more general parameterization of amplitudes for 
$\jpsi \to PP$ decays~\cite{seiden}. With this form, one could easily 
derive the relative decay strength between different final states. 
However, it has also been noticed that under the most general 
circumstances, symmetry breaking introduces too many parameters for 
a general analysis to be useful. In order to reduce the number of 
new parameters considerably and make the analysis manageable, 
only special cases could be considered.
  
\btbl[htb]
\caption{\label{fk_preds}Comparison of predictions~\cite{feldman} and 
measurements~\cite{pdg06} for the branching ratios ($10^{-5}$) for 
$\psp$ decays. The upper limits are presented at 90\% C.L.} 
\vskip 0.2 cm 
\btbu{ccc} \hline \hline 
      VP       & Prediction  &    Measurement        \\ \hline\hline 
   $\rpi$      & $1.3$       &  $ 3.2 \pm 1.2 $      \\
$\KKSN +c.c.$  & $5.1$       &  $10.9 \pm 2.0 $      \\
$\KKSC +c.c.$  & $1.3$       &  $ 1.7^{+0.8}_{-0.7}$ \\
$\omega \eta$  & $1.2$       &  $<1.1$               \\ 
$\omega \etap$ & $6.3$       &  $ 3.2^{+2.5}_{-2.1}$ \\
$\phi \eta$    & $1.6$       &  $ 2.8^{+1.0}_{-0.8}$ \\ 
$\phi \etap$   & $4.6$       &  $ 3.1 \pm 1.6 $      \\
$\rho \eta$    & $2.1$       &  $ 2.2 \pm 0.6 $      \\
$\rho \etap$   & $1.2$       &  $1.9^{+1.7}_{-1.6}$  \\
$\omega \piz$  & $3.8$       &  $ 2.1 \pm 0.6 $      \\
$\phi \piz$    & $0.01$      &  $<0.40$              \\ 
\hline\hline 
\etbu
\etbl

  A similar parameterization with mixing feature of 
the strong interaction mechanism was proposed by Feldmann 
and Kroll (FK)~\cite{feldman} for the
hadron-helicity non-conserving $\jpsi$ and $\psp$ decays, but with a 
different interpretation from those put forth in 
Refs.~\cite{chen,bramon,seiden,tuan}.   
FK assumed that with a small probability, 
the charmonium possesses Fock components built from light quarks only. 
Through these Fock components the charmonium state decays by 
a soft mechanism which is modeled by $\jpsi$-$\omega$-$\phi$ mixing and 
subsequent $\omega$ (or $\phi$) decays into the $VP$ state. 
In absence of the leading-twist perturbative QCD
contribution, the dominant mechanism is supplemented 
by the electromagnetic decay contribution and DOZI violating contribution. 
FK argued that this mechanism can probe the charmonium wave function 
at all quark-antiquark separations and feels the 
difference between a $1S$ and a $2S$ radial wave functions. 
The node in the latter is supposed to lead
to a strong suppression of the mechanism in $\psp$ decays.
With a few parameters adjusted to the experimental data, 
FK obtained a numerical 
description of the branching fractions for many VP decay modes of the 
$\jpsi$ and $\psp$, which agree with the measured branching fractions
at the order of magnitude level, as shown in Table~\ref{fk_preds}. 
Moreover, FK has extended their mixing 
approach to the $\etac\to VV$ decays and obtained a reasonable description
of the branching fractions for these decays while the $\etacp \to VV$ 
decays are expected to be strongly suppressed.

\subsection{Remarks}

  From the brief retrospect of $\RP$ puzzle history, we could notice that
in the early stage, theorists concentrate on peculiar mechanism for
special channels like $\RP$ and $\kstk$, such as Hou and Soni's explanation.
With the development of the theory and the progress of the experiment,
theorists attempt to provide a more general scheme for charmonium decays,
such as the work by Feldmann and Kroll. In fact, the charmonium decay is an 
interconnected system as a whole, a correct explanation of $\rhopi$ puzzle 
is expected to describe, quantitatively, or semi-quantitatively,  
the properties of all measured decay modes.

At last, a few words about Fock state component. 
This concept has been discussed in many works, such as by 
BK~\cite{brokar}, CB~\cite{chen}, and FK~\cite{feldman}.
Even from the theoretical explanation of experiments other than 
$\EE$ collision, the non-perturbative Fock component is also 
indicated to exist. For example, the analysis of charmonia 
photoproduction amplitude implies~\cite{hoyer} 
that the $|\ccb \rangle$ component is narrowly distributed in the 
transversal direction, while the $\ccb$ pairs at larger separations 
may be part of higher Fock states which contain gluons and light quarks.

\section{Evaluation of ratio of $\psp$ to $\jpsi$ decay}\label{sect_rule12}

In this section, we depict three approaches to estimate the ratio of
$\psp$ to $\jpsi$ decay, one from theoretical analysis~\cite{appelquist},
the other two from experimental evaluation~\cite{guyf}.

\subsection{Theoretical Method}\label{ssct_pQCDrule}

   Conventionally, the measured ratios of $\psp$ to $\jpsi$ branching
fractions for specific exclusive hadronic decays are compared with 
the naive prediction of pQCD, the so-called ``12\% rule''.
In the framework of pQCD~\cite{appelquist}, 
the $\psi$ particles are considered to be nonrelativistic bound states
of a charm quark and its antiquark. Their decays into light hadrons are 
believed to be dominated by the annihilation of the $\ccb$ pair into three
gluons. In order to annihilate, the $c$ and $\overline{c}$ must have a
separation of order $1/m_c$, which is much smaller than the size of the 
charmonium state. Thus the annihilation amplitude for an $S$-wave state
like $\jpsi$ and $\psp$ must be proportional to the wave function at the 
origin, $\psi(\mathbf{r}=0)$~\cite{appelquist}. The width for the decay into 
any specific final state $h$ consisting of light hadrons is therefore 
proportional to $|\psi (0)|^2$. The width for the decay into $\EE$ is also 
proportional to $|\psi (0)|^2$. This leads to the simple prediction 
that the ratio of the branching fractions between $\psp$ and $\jpsi$ is
given by Eq.~\eref{qcdrule}.

However, this naive prediction suffers from several apparent approximations. 
Higher order corrections, which may not even be small, are not included 
in this calculation. For example, a first order correction to the branching fraction of 
$\jpsi\to \EE $ could be 50$\%$ of the lowest term if 
$\alpha_{s}(m_{J/\psi})\sim 0.2$~\cite{kwong} is taken. 
The relativistic effect is also ignored. Since the mass difference between $\jpsi$ and 
$\psp$ is around $20\%$ and $<v^{2}/c^{2}>\sim0.24$ for $\jpsi$, this 
correction may be at the same level as the lowest order contribution~\cite{kwong}. 
The inclusion of the finite size of the decay vertex will significantly 
reduce the $ggg$ decay width of the $J/\psi$~\cite{chiang}. Moreover, the 
effect of non-perturbative dynamics is neglected, the size of which is 
hard to estimate. Therefore, people may question the validity of the 
12 $\%$ rule as a serious benchmark for comparing experimental data. 

\subsection{Experiment Estimation (I)}

The first experimental estimation is 
based on the assumption that the decays of $\jpsi$ and $\psp$ in the 
lowest order of QCD are classified into hadronic decays ($ggg$), 
electromagnetic decays ($\gamma^*$), radiative decays into light hadrons 
($\gamma gg$), and transition to lower mass charmonium states 
($\ccb X$)~\cite{kopke,guyf}. Thus, using the relation 
$\BR(ggg)+\BR(\gamma gg)+\BR(\gamma^*)+\BR(\ccb X)=1$, one can derive
$\BR(ggg)+\BR(\gamma gg)$ by subtracting $\BR(\gamma^*)$ and
$\BR(\ccb X)$ from unity.

 \btbl[htb]
\caption{\label{esdat}Experimental data on the branching fractions
for $\jpsi$ and $\psp$ decays through virtual photon and to lower
mass charmonium states used in this analysis. Most of the data are
taken from PDG~\cite{pdg06}, except for $\BR (\jpsi,\psp \ra \gamma^* 
\ra \mbox{hadrons})$, which are calculated by the product $R
\cdot \BR (\jpsi,\psp \ra \MM)$, with $R=2.28\pm 0.04$~\cite{seth}. 
In estimating the errors of the sums, the correlations between the 
channels are considered~\cite{moxh}.} \vskip 0.2 cm \center
\btbu{lcc}
\hline \hline Channel & $\BR(\jpsi)$ & $\BR(\psp) $        \\
\hline \hline $\gamma^* \ra \mbox{hadrons}$
        & (13.50$\pm$0.30)\%& (1.66$\pm$0.18)\%  \\
$\EE$   & (5.94$\pm$0.06)\% & (7.35$\pm$0.18)$\times 10^{-3}$ \\
$\MM$   & (5.93$\pm$0.06)\% &  (7.3 $\pm$0.8)$\times 10^{-3}$ \\
$\TT$   &     $-$           &  (2.8 $\pm$0.7)$\times 10^{-3}$ \\
\hline $\gamma^* \ra X$
        &(25.37$\pm$0.35)\% & (3.41$\pm$0.27)\%  \\ \hline
$\gamma\eta_c$
        & (1.3$\pm$0.4)\%   & (2.6$\pm$0.4)$\times 10^{-3}$ \\
$\PPJP$ &                   & (31.8 $\pm$0.6)\%   \\
$\piz\piz\jpsi$ &           & (16.46$\pm$0.35)\%   \\
$\eta\jpsi$     &           & (3.09 $\pm$0.08)\%  \\
$\piz\jpsi$     &           & (1.26$\pm$0.13)$\times 10^{-3}$ \\
$\gamma\chi_{c0}$&          & (9.2$\pm$0.4)\%    \\
$\gamma\chi_{c1}$&          & (8.7$\pm$0.4)\%    \\
$\gamma\chi_{c2}$&          & (8.1$\pm$0.4)\%    \\  \hline 
$\ccb X$& (1.3$\pm$0.4)\%   &(77.7$\pm$1.2)\%    \\ \hline \hline \etbu
\etbl

The calculated values of $\BR(\gamma^*)$ and $\BR(\ccb X)$, together with the values 
used to calculate them are summarized in Table~\ref{esdat}. As regards to $\psp$, 
two final states $\gamma \eta (2S)$ and $h_c(1^1P_1)+X$ with faint branching fractions 
are neglected in our calculation. By deducting the contributions $\BR(\gamma^*)$ and 
$\BR(\ccb X)$, we find that $\BR(\jpsi \ra ggg)+\BR(\jpsi \ra \gamma gg)
= (73.3\pm 0.5)\%$ and $\BR(\psp \ra ggg)+\BR(\psp \ra \gamma gg)=
(18.9\pm 1.3)\%$, then the ratio of them is 
 \beq Q_g =\frac{\BR(\psp \ra ggg + \gamma gg)}
               {\BR(\jpsi \ra ggg + \gamma gg)} = (25.7 \pm 1.7)\%~.
 \label{qvone} \eeq
The above estimation is consistent with the previous
ones~\cite{suzukic,guyf}. The relation between the decay rates of
$ggg$ and $\gamma gg$ is readily calculated in pQCD to the first
order as~\cite{kwong}
 \begin{equation*} 
\frac{\Gamma(\jpsi \ra \gamma gg)}{\Gamma(\jpsi \ra ggg)}
 = \frac{16}{5} \frac{\alpha}{\alpha_s(m_c)}
 \left( 1- 2.9 \frac{\alpha_s}{\pi} \right). 
\end{equation*}
Using $\alpha_s(m_c)=0.28$, one can estimate the ratio to be
0.062. A similar relation can be deduced for the $\psp$ decays.
So we obtain $\BR(\jpsi \ra ggg) \simeq (69.0\pm0.5)\%$ and
$\BR(\psp \ra ggg) \simeq (17.8\pm 1.2)\%$, while the ``25.7\%
ratio'' in Eq.~(\ref{qvone}) stands well for both $ggg$ and
$\gamma gg$. Although $Q_g$ is considerably enhanced relative to
$Q_h$ in Eq.~(\ref{qcdrule}), it is fairly compatible with the ratios 
for the $\kk$ and $\kskl$ decay modes between $\psp$ and $\jpsi$, which are
\[\begin{array}{rcl}
Q_{\kk}   &=& (26.6 \pm 4.5)\%~\mbox{(CLEO)}~, \\
Q_{\kskl} &=& (28.8 \pm 3.7)\%~\mbox{(BES)}~,\\
Q_{\kskl} &=& (32.2 \pm 5.2)\%~\mbox{(CLEO)}~,\\
\end{array}\]
according to the recent results from CLEO and 
BES~\cite{cleocpp,jpsikskl,psipkskl}. The relation in
Eq.~(\ref{qvone}) was discussed in the literature as the hadronic
excess in $\psp$ decays~\cite{guyf,suzukic}. It implicates that
while some modes are suppressed in $\psp$ decays, the dominant part of
$\psp$ through $ggg$ decays is enhanced relative to the 12\% rule 
prediction in the light of $\jpsi$ decays.

\subsection{Experiment Estimation (II)}

The second approach for estimating $Q_h$ is to use the data on branching 
fractions for hadronic decays in final states containing pions, kaons, 
and protons that have already been measured for both the $\jpsi$ and 
the $\psp$. They are $\pp$, $\kk$, $\ppb$, $\pp\piz$, $\ppb\piz$, 
$2(\pp)$, $3(\pp)$, $2(\pp)\piz$, $3(\pp)\piz$, $2(\pp\piz)$, $2(\kk)$, 
$\kk\pp$, $\kk\pp\piz$, $\kk2(\pp)$, $\pp\ppb$, $\ppb\pp\piz$ and so forth. 
Using the data compiled in Table~\ref{qval}, we have
$$ \sum\limits^{11}_{i=1} \BR (\jpsi \ra f_i) =(14.46 \pm 0.80)\%~ $$
and
$$ \sum\limits^{11}_{i=1} \BR (\psp \ra f_i) =(1.70 \pm 0.23)\%~. $$
It follows that
\beq
\begin{array}{ccl}
 Q_s &=& \sum\limits^{11}_{i=1} \BR (\psp \ra f_i)  \bigg/
         \sum\limits^{11}_{i=1} \BR (\jpsi \ra f_i)  \\
     &=& (11.8 \pm 1.7 )\%~.
\end{array}     
\label{qvtwo}
\eeq

\begin{table}[bth]
\caption{\label{qval}Branching fractions for $\psp$ and 
$\jpsi$ decays, and $Q_h$ values are also calculated. 
For $\psp\to\pp$, $\kk$ decays, the results are the weighted average of
measurements from CLEO~\cite{cleocpp} and BES~\cite{yesw}, while for 
$\jpsi\to\pp\piz$, the result is the weighted average of measurements
from BES~\cite{bes3pi} and BABAR~\cite{babar3pi}.
Except for values with $\dagger$, all others from PDG~\cite{pdg06}.}
\center
{
\begin{tabular}{cccc} \hline \hline
  Final state    &${\cal B}_{\jpsi} (10^{-3})$
                                  &${\cal B}_{\psip} (10^{-4})$
                                                    &$Q_h$ (\%) \\ \hline
$\pp$   &$0.147\pm0.023$  &$0.08\pm 0.05\dagger$  & $ 5.6\pm  3.5$  \\
$\kk$   &$0.237\pm0.031$  &$0.63\pm 0.07\dagger$  & $26.5\pm  4.5$  \\
$\ppb$  & $2.17\pm 0.08$  &$2.65\pm 0.22$         & $12.2\pm  1.1$  \\
$\pip\pim\piz$                                    
  &$21.2\pm 1.01\dagger$  &$1.68\pm 0.26$         & $0.79\pm 0.13$  \\
$\ppb\piz$                                        
        & $1.09\pm 0.09$  &$1.33\pm 0.17$         & $12.2\pm  1.9$  \\
$2(\pp)$& $3.55\pm 0.23$  &$2.4 \pm 0.6 $         & $ 6.8\pm  1.8$  \\   
$3(\pp)$& $4.3 \pm 0.4 $  &$3.5 \pm 2.0 $         & $ 8.1\pm  4.7$  \\   
$3(\pp)\piz$                                      
        & $29  \pm   6 $  &$35  \pm 16  $         & $12.1\pm  6.1$  \\
$2(\pp)\piz$                                      
        & $33.7\pm 2.6 $  &$26.6\pm 2.9 $         & $ 7.9\pm  1.1$  \\
$2(\pip\pim\piz)$                                 
        & $16.2\pm 2.1 $  &$45  \pm 14  $         & $27.8\pm  9.4$  \\
$2(\kk)$& $0.78\pm 0.14$  &$0.60\pm 0.14$         & $ 7.7\pm  2.3$  \\
$\kk\pp$& $7.2 \pm 2.3 $  &$7.2 \pm 0.5 $         & $10.0\pm  3.3$  \\
$\kk\pip\pim\piz$                                 
        & $12.0\pm 3.0 $  &$12.4\pm 1.0 $         & $10.3\pm  2.7$  \\
$\kk 2(\pp)$                                      
        & $4.7 \pm 0.7 $  &$18  \pm 9   $         & $38.3\pm 20.0$  \\
$\ppb\pp$                                         
        & $6.0 \pm 0.5 $  &$6.0 \pm 0.4 $         & $10.0\pm  1.1$  \\
$\ppb\pip\pim\piz$        
        & $2.3 \pm 0.9 $  &$ 7.3\pm 0.7 $         & $31.7\pm 12.8$  \\
	                                                \hline\hline
\end{tabular}  }
\end{table}

Some remarks are in order here. First, we know that most of the multihadron final 
states in fact include sums of several two-body intermediate states, so the 
$Q_s$ is not the exact ratio of $\psp$ to $\jpsi$ inclusive hadronic decay rates,
but the ratio on average of the exclusive decay channels as measured to date.
In another word, $Q_s$ represents a mixed effect which may deviate noticeably 
from the component $Q$-values. For example, the decay $\psp \ra \pp\kk$ can proceed 
through intermediate state $K^*(892)^0  \overline{K}_2^* (1430)^0 +c.c.$, 
whose $Q=(2.9\pm1.3)\%$, is greatly suppressed~\cite{bes2vt} comparing with 
$Q=(10.0\pm3.3)\%$. Second, we notice that the results obtained by two estimations 
vary considerably; furthermore, by virtue of Table~\ref{qval}, it is obvious that
many $Q$-values deivate from $Q_s$ significantly while on average, $Q_s$ is similar to 
pQCD $Q_h$. It would be an intriguing problem that with more and more data and higher 
accurate measurements if $Q_s$ and $Q_g$ could approximate with each other and be 
consistent with $Q_h$, or if $Q_s$ and $Q_g$ still deviate from each other prominently.

\section{Comment on the $\RP$ puzzle}

\subsection{Theory Aspect}\label{ssct_theap}

  For all explanations involving $\RP$ puzzle, there is a basic assumption
that the non-relativistic potential model delineates the physics of 
charmonium decays to a good approximation. However, this assumption indeed 
requires examination in detail. First, as we have noticed in 
subsection~\ref{ssct_pQCDrule}, several corrections, should be added for 
the decay ratio between $\psp$ and $\jpsi$. 

  Second, we see the effect of non-perturbative dynamics is neglected, 
which is crucial for charmonium decays. Actually from various 
explanations of $\RP$ puzzle, or more generally from the phenomenological 
explanation involving pQCD, we note that certain non-perturbative effect 
or nonlinear effect must be incorporated one way or the other in order to 
recount the experimental data.
But this kind of effect could hardly been
included in the present non-relativistic potential model. 

Third, if the $S$-$D$ wave mixing scenario holds as the solution 
of $\rhopi$ puzzle, then the matrix element of $D$ wave to light 
hadrons would be very large, which can hardly be accommodated in 
the potential model, or any other current theory. It indicates that the 
current understanding of charmonium decays may not be complete. 
  
We envisage that new development of the theory should take into 
account the following features involving charmonium decays

\begin{itemize}
\item the mass effect for different charmonium states, such as
$\jpsi$, $\psp$, $\pspp$ and so on;
\item the non-perturbative or non-linear effect on the resonance 
decays;
\item reasonable description for the known features of charmonium 
spectroscopy;
\item quantitative consistence (with reasonable high accuracy) with 
the existing experimental measurements.
\end{itemize}

   Herein it is also important to distinguish the quarkonium states 
(theoretical states) and mass eigenstates 
(physical states)~\cite{gerard,artoisenet}. 
From a fairly theoretical point of view, if both states, say, 
$\jpsi$ and $\psp$ did dominantly annihilate into three gluons, 
they would mix and could thus not be the putative quarkonium states.
In a non-relativistic potential model, for example, the $\psp$ is simply 
a radial excitation of the $\jpsi$. This is a well defined picture in 
which $\jpsi$ and $\psp$ are orthogonal states. If the annihilation into 
three gluons could be treated as a ``perturbation'' to the 
non-relativistic potential, then clearly the unperturbed states would 
mix and rearrange themselves into orthogonal mass eigenstates.
The QCD dynamics may be such that the physical states, presumably
mixtures of the theoretical $\ccb$ states ($n^{2S+1} L_J$),
are so built up that one of them strongly annihilates into three
perturbative gluons while the other does not. Mixing of the $1^3 S_1$
and $2^3 S_1$ states via three perturbative gluons has little effect on
the charmonium mass spectrum, but may be crucial for the decay pattern.

\subsection{Phenomenology Aspect}

  One may remember that at the early stage of the discovery of a narrow 
state $\jpsi$, the $\ccbar$ system was hailed as the Hydrogen atom 
of Quantum Chromodynamics, with the implied hope that the study of the 
newly discovered system could shed as much light on the dynamics of 
quark-antiquark interactions as the study of the Hydrogen atom had on 
Quantum Electrodynamics. But one may also notice the historical fact, 
even before Bohr's theory, Balmer series had been discovered for long, 
and the famous Rydberg formula had also been proposed, which laid a 
solid foundation for further theoretical improvement. If we are 
conscious of the more complicatedness of charmonia system comparing 
with Hydrogen atom, we may prepare for more hard and meticulous works. 
As a first step, it is necessary to develop a reliable and extensively 
applicable phenomenological model (PM).  

  The advantage of PM lies in that a well-defined PM contains few 
experimentally determined parameters which have clear physical meaning;
moreover, with only few parameters determined from experiment, PM could 
produce concrete results which can be directly confirmed or falsified 
by experiment and may guid experimental searches.
Such a model has a good relation with elementary principle of the theory, 
and if correct, can be used for further theoretical refinement. 
This point is noteworthy for the time being, since the general QCD can 
hardly provide solutions for special problems; conversely, we have to 
establish certain effective empirical model to advance our understanding 
for generic QCD principle.  

  Here we would like to mention few ideas of PM, which have or intend to 
provide a general description for charmonium decay. 

\subsubsection{Mixing Model}

As we have noticed in subsection~\ref{sst_vmmm}, whatever CI model or
FK model, they could yield definite predictions for experimental test, 
and therefore provide clues for further development. In addition, 
according to Haber and Perrier's parameterization scheme~\cite{haber},
we could get the decay rate relations between different channels, and the
proportions of different interaction amplitudes, all of which are 
useful information to understand the dynamics of charmonium decays.

The special feature of Rosner's $S$- and $D$-wave mixing 
scheme~\cite{rosnersd} is that it is simple, and it works both for 
suppressed and enhanced decay modes, and moreover, it connects 
the $\jpsi$, $\psp$ and $\pspp$ decays together, and 
give predictions with little uncertainty. 

\subsubsection{Effective FSI}
Miller once discussed the connection between the strong-coupling
approximation to quantum chromodynamics and nuclear properties
observed at low and medium energies and momentum transfer~\cite{miller}. 
He suggested that the strong-coupling (corresponds to long-range 
interaction) version of QCD does reproduce the salient feature of 
the meson-baryon picture of low momentum transfer in nuclear physics. 
The derivation of nuclear physics indicates that quark aspects of 
ordinary nuclei are hidden in the hadronic degrees of freedom. 
Reversely, one may imagine that a quasi-meson or quasi-baryon 
structure could be formed in non-perturbative hadronization process.
The meson-baryon picture in nuclear physics could be utilized, with 
some modification, as an effective FSI theory, or molecular-model theory. 
The recent molecular interpretations~\cite{liux,wymcxog} for the newly 
found state $Y(4260)$ could be treated as such an effective FSI theory. 

Furthermore, the quasi-particles could even be real particles, and then
with the residual strong force between the quarks inside the 
quasi-particles, some multiplets can be formed as the quasi-particle and
anti-quasi-particle bound states. This idea was first put forth to
explain a lots of low mass baryon-antibayron enhancements newly found
by CLEO, Belle and BaBar collaborations~\cite{wymbbn}.

\subsubsection{Glueball and Hybrid}
We have seen that the $\jpsi$-glueball mixing scheme is the first explanation 
proposed to explain the $\rpi$ puzzle. With the implication from lattice
calculation~\cite{morningstar}, Suzuki once proposed a glueball-$\psp$ 
mixing scheme to explain the excess hadronic decays at 
the $\psp$~\cite{suzukid}. In fact, gluodynamics is always a tantalizing 
domain for theorists. Recently, some lattice evaluations 
suggest~\cite{chiutw,luoxq} that the masses of some hybrid states could 
be low enough to be degenerated with charmonium states, such as $\jpsi$ 
and/or $\chicj$. Unlike Glueball, these $\ccb g$ hybrid states could be 
very narrow, so it may be very stimulating and noteworthy to search for 
and confirm these kinds of states.

\subsection{Experiment Aspect}

  Physics is a science of experiment. Physical facts are bases for
theoretical development and also criteria for checking theoretical
hypotheses. It is a prominent fact that the early analyses on $\RP$ puzzle 
based on meager experimental data often lead to unsatisfactory, sometimes 
premature, inferences, which were washed out easily by later accurate data. 
In fact, the current knowledge concerning the $\psp$ decays from experiments
is still rather limited, even summing all charmless channels presented
in PDG~\cite{pdg06}, the total branching fraction is less 2\%. Such 
situation prevents us from laying down a solid foundation for elementary 
dynamics exploration. However, the estimation of $Q$ value, as discussed 
in Section~\ref{sect_rule12}, affords us some clues concerning the 
exploration of charmonium decay dynamics. Since many suppressed channels 
have been found, especially those such as $\rpi$ which is greatly 
suppressed in $\psp$ decay, and if the $Q_h$ really represents the 
averaged value of inclusive hadronic decay, the estimation of $Q_g$ 
indicates that either lots of enhanced decays are not discovered, or 
some particular decays only present in $\psp$, or both cases exist. 
Therefore, systematic experimental study of $\psp$ decays is 
anxiously awaited.

Moreover, study should be carried out not only on the ratios of $\psp$ to 
$\jpsi$ decays, but also on the other ratios such as those between $\etacp$ and 
$\etac$~\cite{feldman}, and those between $\pspp$ and $\jpsi$\cite{yrhopi},
and/or many other ratios between different resonances for the same
channel or between different channels about the same 
resonance~\cite{Chernyak}. All these studies will shed light on the
understanding of charmonium decays. 
From our point of view, the progress would be more likely obtained from 
the analysis of the new experimental results, rather than from the inspiration 
of general theoretical principle.
   
\section{Summary}

In this paper, we present a general review on the study of the $\RP$ puzzle. 
Although there is still no satisfactory explanation for all existing 
experiment results, some enlightenment ideas have been put forth.
In addition, we also discuss three methods of estimating the ratio 
of the branching fractions in $\jpsi$ and $\psp$ decays. In the light of 
the present theoretical and experimental status, we argue that it is 
important to explore the potential models from a new point of view, 
it is necessary to search for and/or construct an effective 
phenomenological model, and it is especially crucial to perform systematic 
measurements of various charmonium decays.

As we know physics in the charm energy region is in the boundary 
domain between perturbative and nonperturbative QCD. 
Recently the observed
hadronic decays of charmonium may give new challenges to the present 
theoretical understanding of the decay mechanisms. 
Many of the new charmonium states observed by Belle and BaBar
which can hardly be accommodated by potential models 
have led to new theoretical
speculations about charmonium spectroscopy and novel production
mechanism~\cite{Swanson}.

Experimentally, the expected large data sample from CLEOc in the near future, 
and even larger sample from forthcoming detector BESIII will open for us 
a new era for charmonium dynamics study, even though we may not obtain 
a throughly revolutionary theory, we could acquire more profound 
understanding for the existing theory, at the same time we could expect
a brand-new comprehension for the $\rpi$ puzzle.

\acknowledgments

This work is supported in part by the National Natural Science Foundation
of China under Contract No. 10491303 and the 100 Talents Program of CAS
under Contract No. U-25.

\end{document}